\documentclass[a4paper,11pt]{article}
\usepackage{pos}

\title{Heavy flavor and jet studies for the future Electron-Ion Collider}

\author*[a]{Xuan Li}

\affiliation[a]{Physics Division, Los Alamos National Laboratory,\\
  MS H846, Los Alamos, NM, 87545, USA}

\emailAdd{xuanli@lanl.gov}

\abstract{The proposed high luminosity high energy Electron-Ion Collider (EIC) will explore the proton/nuclear structure in a wide Bjorken-x ($x_{BJ}$) and $Q^{2}$ phase space. Heavy flavor products are generated in initial collisions and have their hadronization influenced by the nuclear medium. Heavy flavor hadron and jet measurements at the future EIC will allow us to better constrain the nPDFs especially in the poorly constrained high $x_{BJ}$ region, precisely determine the quark/gluon fragmentation processes and directly study the quark/gluon energy loss within the nuclear medium. These measurements can constrain the cold nuclear medium effects for previous and ongoing heavy ion measurements at the Relativistic Heavy Ion Collider (RHIC) and the Large Hadron Collider (LHC). Silicon vertex/tracking detectors are essential to realize the heavy flavor and jet measurements at the EIC. Feasibility studies in simulation for proposed heavy flavor and jet observable with evaluated detector performance at the future EIC will be discussed.}

\FullConference{%
  HardProbes2020\\
  1-6 June 2020\\
  Austin, Texas}

\begin{document}
\maketitle

\section{Introduction}
The proposed Electron-Ion Collider (EIC) in US will utilize high energy high luminosity Deeply Inelastic Scattering (DIS) processes in $e+p$ and $e+A$ collisions to explore fundamental questions in nuclear physics such as the three dimensional partonic structure of nucleons and nuclei \cite{ref_white_paper}. The EIC has reached the CD0 stage and will be built at the Brookhaven National Laboratory. It will make electron+nucleus collisions with multiple nuclear species ranging from $^{3}\mathrm{He}$ to $^{208}\mathrm{Pb}$ at a series of center-of-mass energies from 20 GeV to 141 GeV. A clean environment will be provided by the future EIC to study the nuclear medium effects on both initial state nuclear parton distribution functions (nPDF) and final state hadronization processes. 

Heavy flavor quarks are produced in initial collisions due to their heavy masses ($m_{c/b} >> \Lambda_{QCD}$). They serve as an ideal probe to access the impact of the nuclear medium on initial-state and final-state observables. Unlike heavy ion experiments at Relativistic Heavy Ion Collider (RHIC) and the Large Hadron Collider (LHC), these asymmetric collisions at the EIC have no multiple interactions of the electron with the nucleus. Therefore heavy flavor production at the EIC can precisely study the initial state nPDFs especially in the large $x_{BJ}$ region \cite{npdf1, npdf2} and the final state fragmentation/hadronization processes \cite{RefB}. The polarized electron+nucleon/nuclei collisions at the future EIC will provide opportunities to further explore the nucleon/nucleus spin structure such as the gluon Sivers effects through heavy flavor measurements as well. A physics program, which focuses on the EIC heavy flavor and jet physics developments together with the related detector conceptual design and R$\&$D, has been established at Los Alamos National Laboratory (LANL) with Laboratory Directed Research and Development (LDRD) funding supports \cite{lanl_eic}. Initial results on heavy flavor hadron and jet studies in simulation associated with this LANL EIC project will be discussed.

\begin{figure*}[!ht]
\begin{center}
\includegraphics[width=0.84\linewidth]{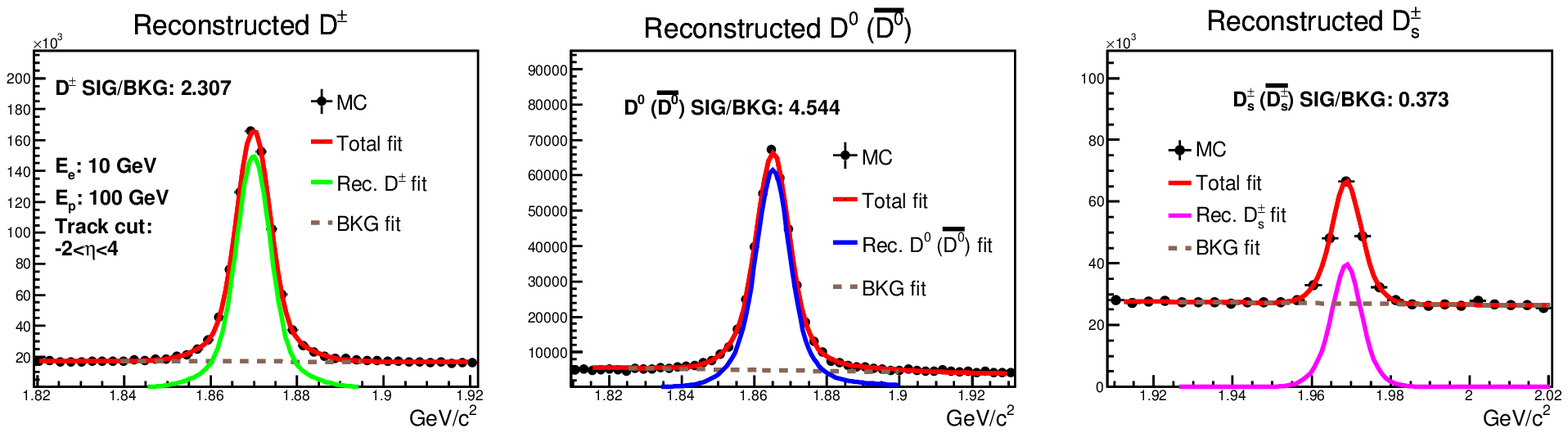}
\includegraphics[width=0.56\linewidth]{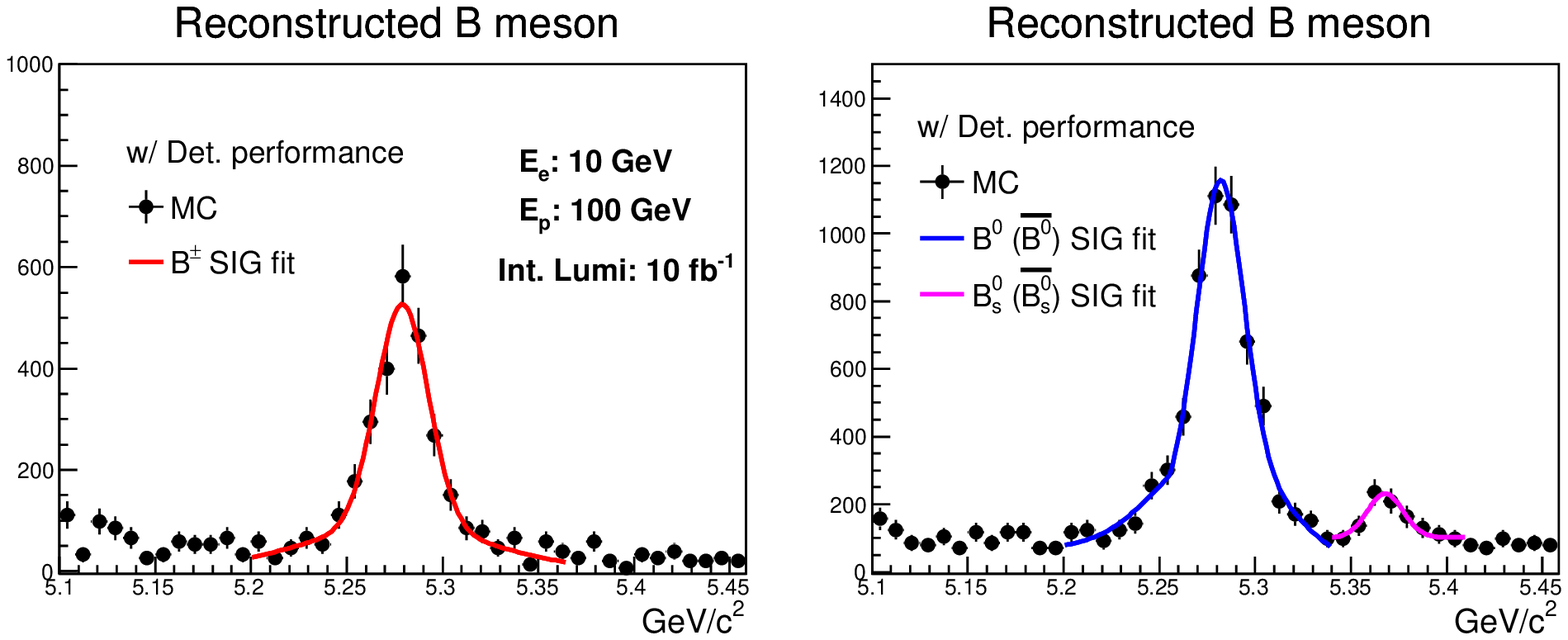}
\caption{\label{fig:hf_had_reco} Simulation studies of reconstructed D-meson and B-meson mass spectrum in 10 GeV electron and 100 GeV proton collisions with integrated luminosity at 10 $fb^{-1}$. Good signal over combinatorial background ratios have been achieved for reconstructed $D^{\pm}$, $D^{0}$ ($\bar{D^{0}}$) and $D_{s}^{\pm}$. Clear $B^{\pm}$, $B^{0}$ ($\bar{B^{0}}$) and $B_{s}^{0}$ and ($\bar{B_{s}^{0}}$) signals have been obtained.}
\end{center}
\end{figure*}

\section{Heavy flavor hadron studies in simulation}
The full analysis framework for EIC heavy flavor hadron and jet studies has been established, which includes event generation in PYTHIA \cite{py8}, implementation of evaluated detector (vertex, particle identification and tracking) performance and hadron/jet reconstruction algorithm developments. The heavy flavor production at the future EIC is dominated by the neutral current ($\gamma^{*} / Z$ exchange) process in DIS. The detector response contains 20 $\mu$m primary vertex $x-y$ resolution, $95\%$ $\pi/K/p$ identification efficiency over the entire acceptance, and proposed central+forward (nucleon/nuclei beam going direction) silicon vertex/tracking detector performance. Several advanced silicon technologies are considered for the EIC silicon vertex/tracking detector conceptual design. In this simulation, we use silicon sensors with pixel pitch at 30 $\mu$m, average material budgets per layer at $0.4 \% X_{0}$ and 500 kHZ readout rate as the detector reference. A hybrid design of five barrel layers and five forward planes with various technologies for the EIC vertex/tracking detector is under study. The collision set of 10 GeV electron and 100 GeV proton is used to study the fully reconstructed heavy flavor hadrons. To search for D-meson and B-meson candidates, charged tracks are clustered by matching their transverse decay length within 90 $\mu$m.

\begin{figure*}[ht]
\begin{center}
	\includegraphics[width=0.41\linewidth]{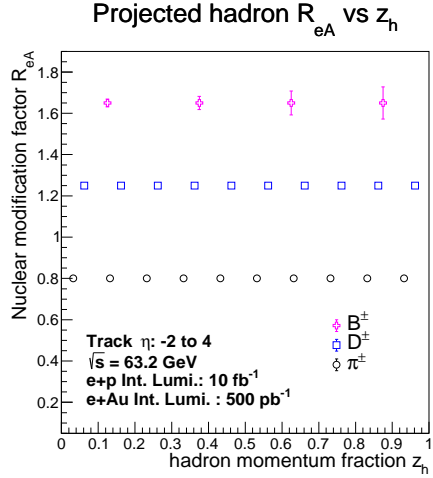}
	\caption{\label{fig:rea_had} Projected nuclear modification factor $R_{eAu}$ for reconstructed hadrons versus hadron momentum fraction $z_{h}$ with statistical uncertainties. The statistical uncertainties are determined in 10 $fb^{-1}$ $e+p$ collisions and 500 $pb^{-1}$ $e+Au$ collisions. Only charged tracks within pseudorapidity -2 to 4 region are used to hadron reconstruction. Black open circles represent charge pions, blue open rectangular points are for D-mesons and B-mesons are shown in magenta open cross points.}
\end{center}
\end{figure*}

\begin{figure*}[ht]
 	\begin{center}
	\includegraphics[width=0.5\linewidth]{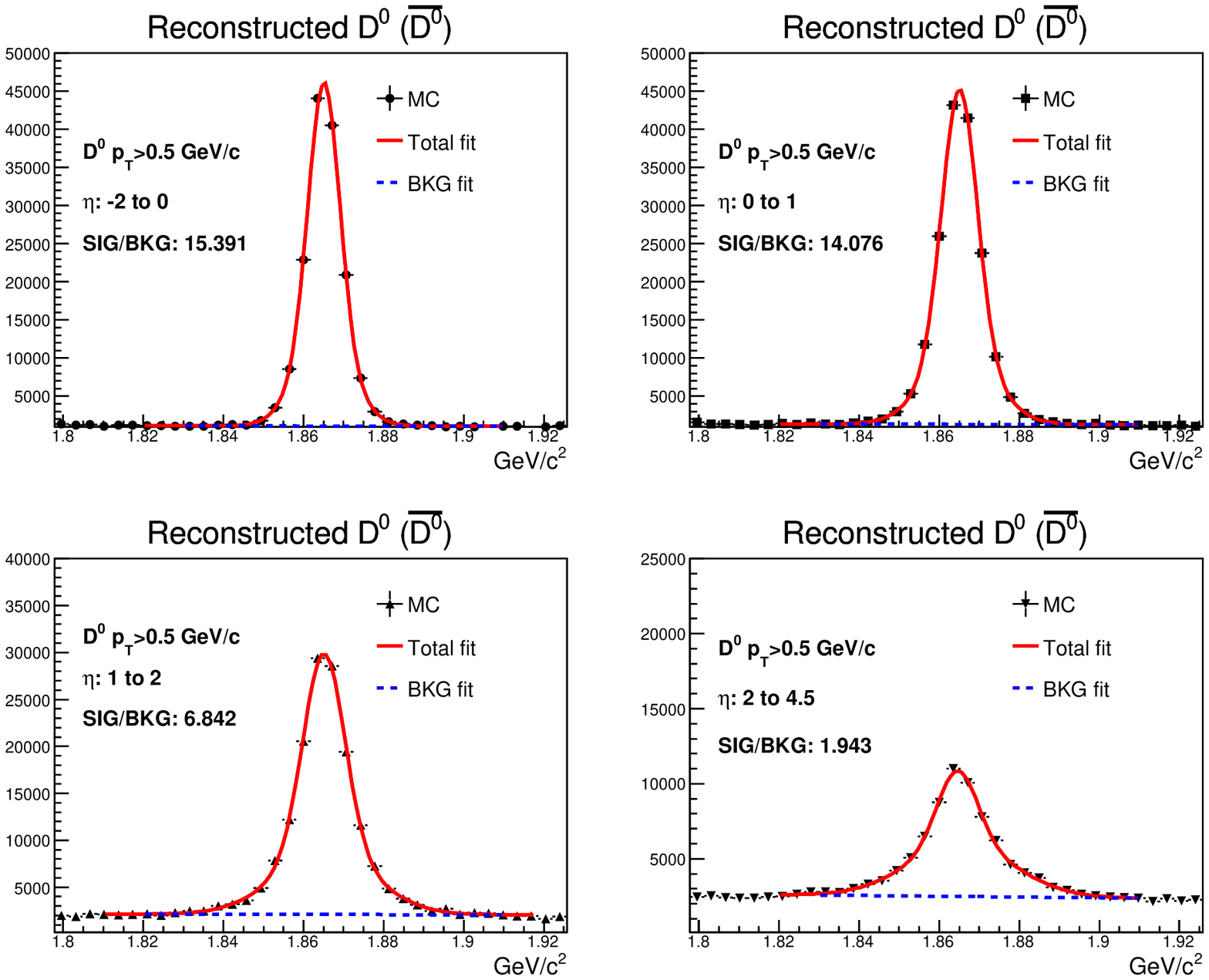}
	\includegraphics[width=0.39\linewidth]{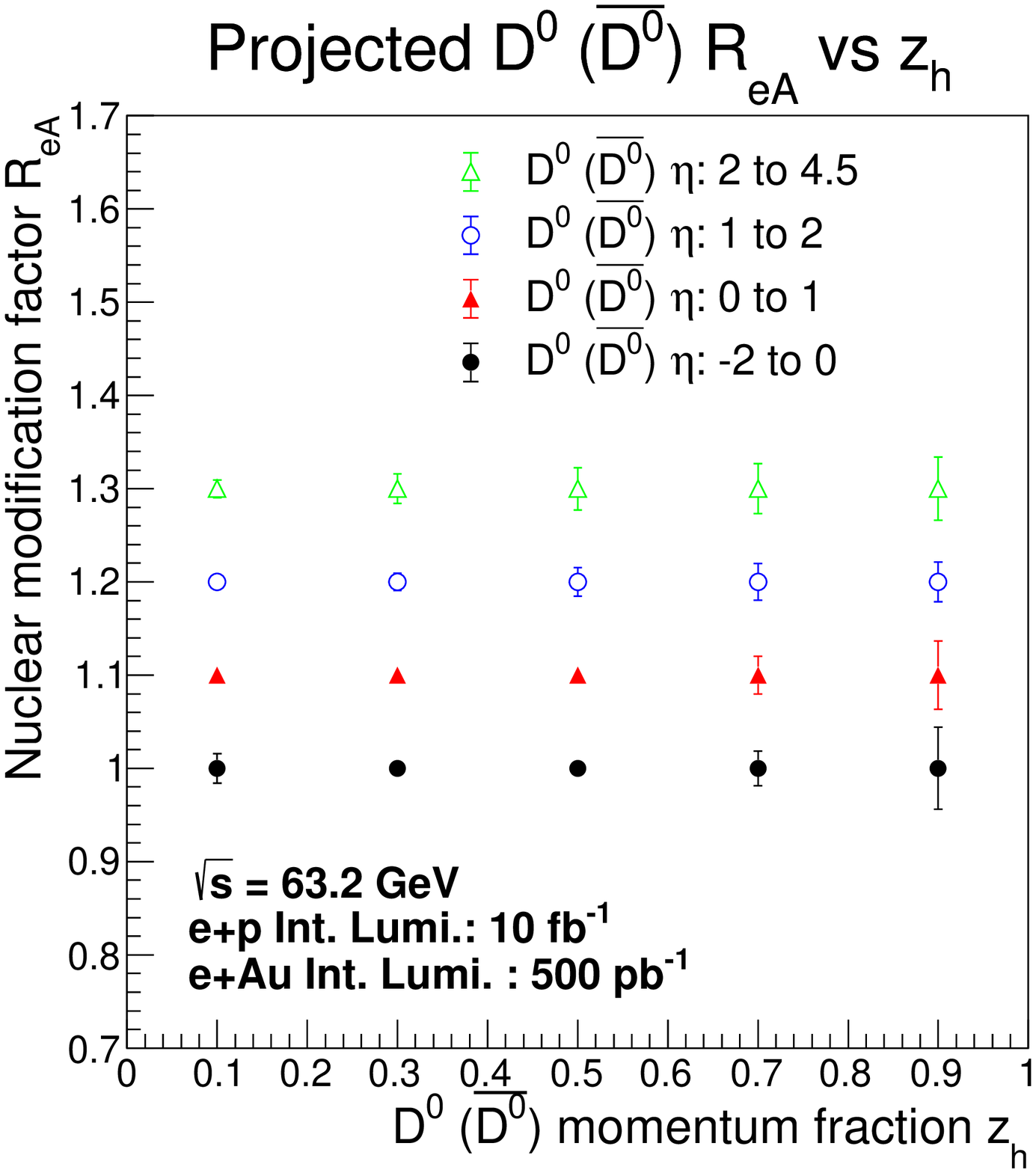}
	\caption{\label{fig:rea_d0} Reconstructed $D^{0}$ ($\bar{D^{0}}$) mass spectrum within different pseudorapidity regions in $e+p$ collisions at $\sqrt{s}$ = 63.2 GeV (left) and projected hadron momentum fraction $z_{h}$ dependent nuclear modification factor $R_{eAu}$ for reconstructed $D^{0}$ ($\bar{D^{0}}$) with associated kinematic regions (right). The statistical uncertainties are determined in 10 $fb^{-1}$ $e+p$ collisions and 500 $pb^{-1}$ $e+Au$ collisions.}
\end{center}
\end{figure*}

Figure~\ref{fig:hf_had_reco} shows the reconstructed D-mesons and B-mesons from these charged track clusters in 10 GeV electron and 100 GeV proton collisions with 10 $fb^{-1}$ integrated luminosity. Good signal over combinatorial background ratios have been achieved for reconstructed $D^{\pm}$, $D^{0}$ ($\bar{D^{0}}$) and $D_{s}^{\pm}$. Fully reconstructed $B^{\pm}$, $B^{0}$ ($\bar{B^{0}}$), and $B_{s}^{0}$ ($\bar{B_{s}^{0}}$) have been obtained as well. The projected nuclear modification factor $R_{eAu}$ for reconstructed hadrons as a function of the hadron momentum fraction $z_{h}$ are shown in Figure~\ref{fig:rea_had}. The mean values are evaluated based on parton energy loss calculations \cite{RefB,lanl_eic}. The statistical uncertainties are determined for reconstructed hadron yields in 10 $fb^{-1}$ $e+p$ collisions and 500 $pb^{-1}$ $e+Au$ collisions. These heavy flavor measurements provide great discriminating power to explore the hadronization process in nuclear medium and help separate different model predictions with good precisions. More forward hadron measurements can access higher $x_{BJ}$ partons, good signal over background ratios and high statistics can be obtained for reconstructed $D^{0}$ ($\bar{D^{0}}$) from backward to forward pseudorapidity regions in the same simulation sample as shown in left side of Figure~\ref{fig:rea_d0}. The corresponding projected statistical uncertainties for reconstructed $D^{0}$ ($\bar{D^{0}}$) in these kinematic regions are shown in the right panel of Figure~\ref{fig:rea_d0}.

\section{Heavy flavor jet tagging and jet substructure studies in simulation}
\begin{figure*}[ht]
 	\begin{center}
	\includegraphics[width=0.32\linewidth]{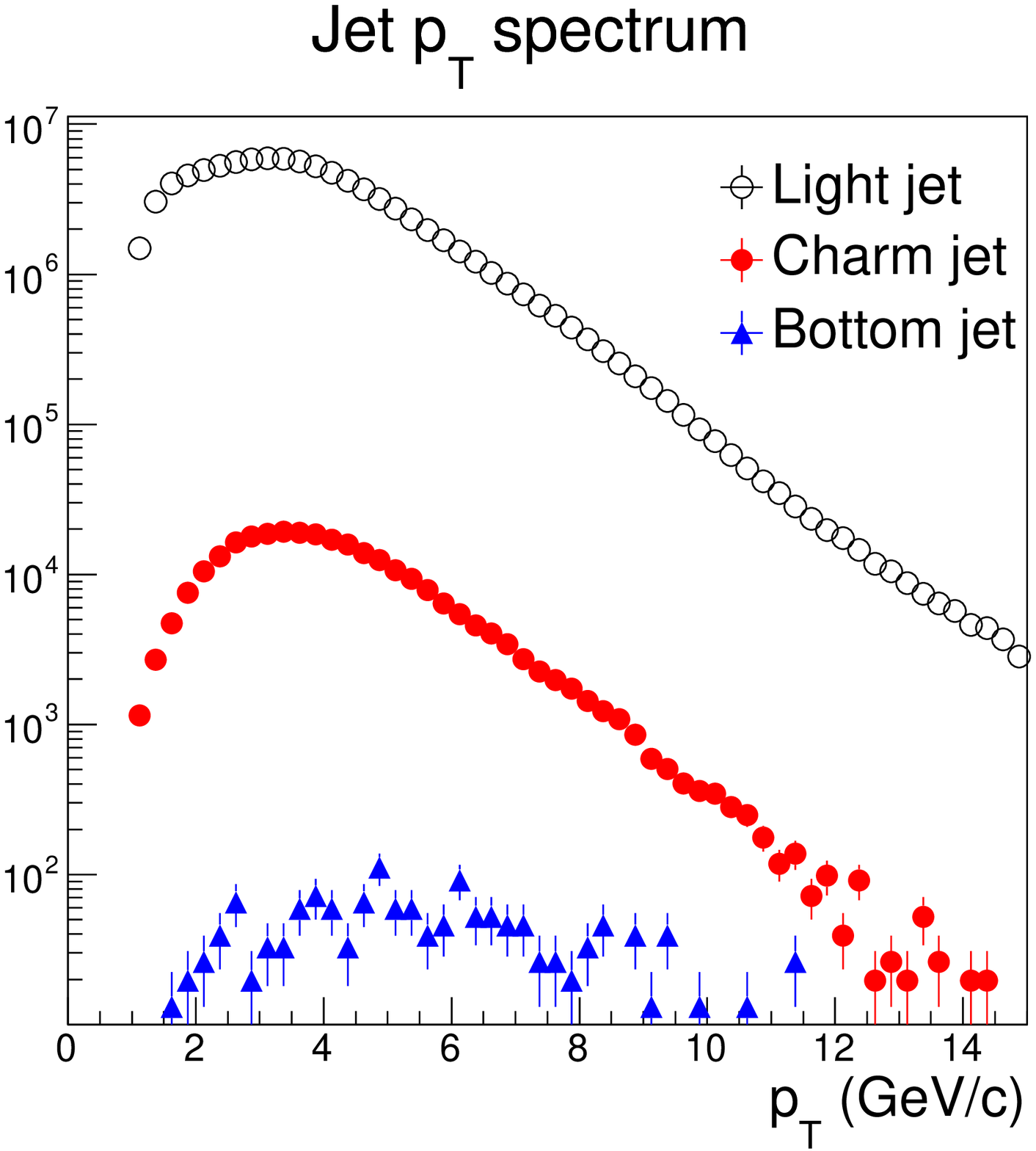}
	\includegraphics[width=0.56\linewidth]{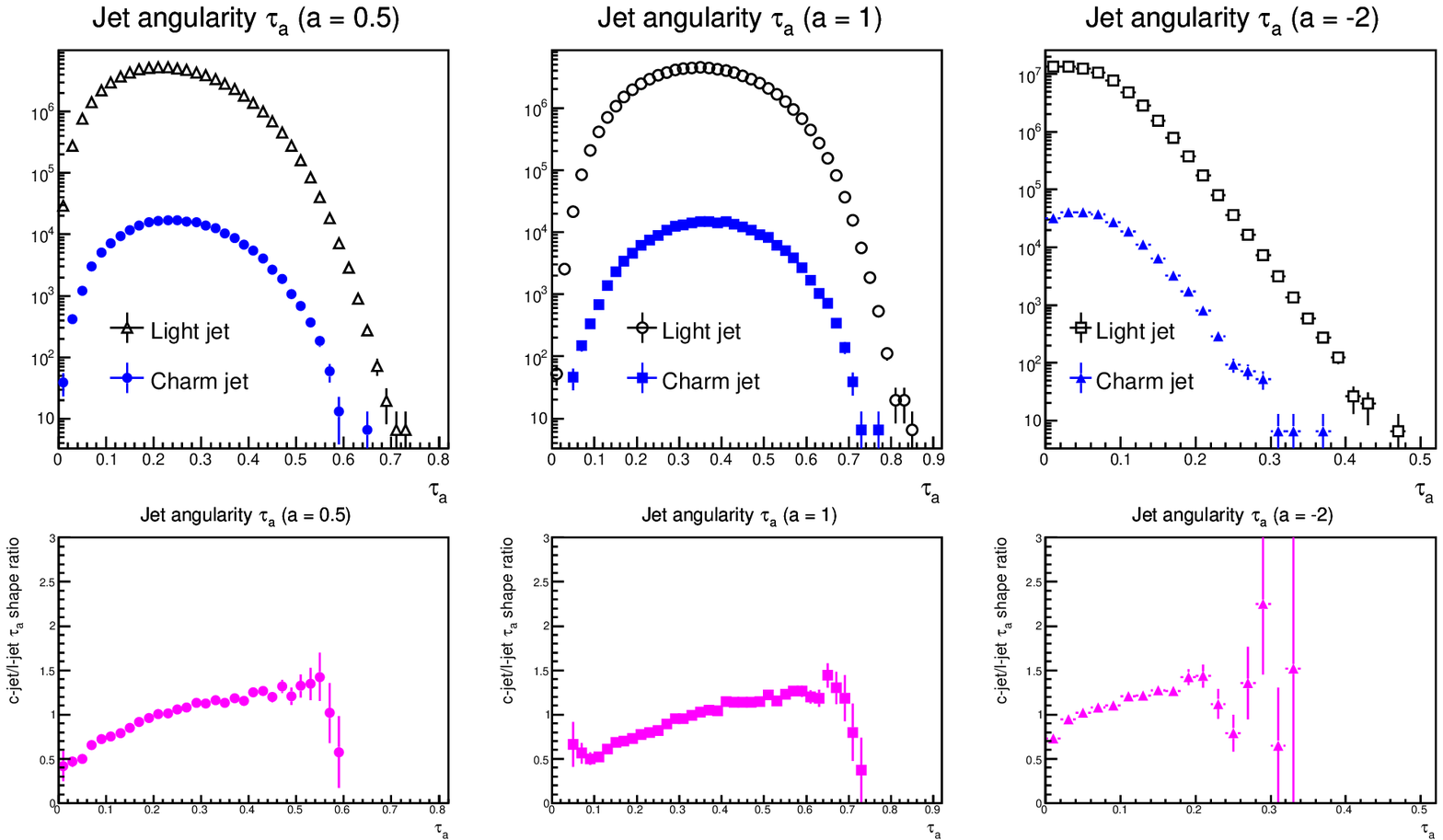}
	\caption{\label{fig:eic_jet} Left panel: Jet $p_{T}$ spectrum for light-flavor jets (black open circles), charm-jets (red closed circles) and bottom-jets (blue closed triangles) in $e+p$ collisions at $\sqrt{s}$ = 63.2 GeV with 10 $fb^{-1}$ integrated luminosity. Right panel: Jet angularity $\tau_{a}$ distributions for light-flavor jets (black open points) and charm-jets (blue closed points) with different $a$ values in the same simulation setup are shown in the top row and ratios between the two normalized distributions are shown in the bottom row.}
\end{center}
\end{figure*}

The future EIC will be a jet factory to directly probe the initial parton kinematics. In this simulation study, jets are reconstructed with the anti-$k_{T}$ algorithm \cite{anti_kt} and the jet cone radius is selected at 1.0 considering the relativistic low hadron multiplicity in $e+p$ and $e+A$ collisions. The four-vector momentum of reconstructed heavy flavor hadrons is used to tag jets. Jets contain at least one fully reconstructed heavy flavor meson inside their cone are defined as heavy-flavor jets. Jets with no associations with reconstructed heavy flavor hadrons are named as light-flavor jets. The left panel of Figure~\ref{fig:eic_jet} shows the $p_{T}$ spectrum of the different flavor tagged jets in 10 $fb^{-1}$ $e+p$ collisions at $\sqrt{s}$ = 63.2 GeV. Jet substructure provides enhanced sensitivity to explore the hadronization processes. As one of the jet substructure observables, the jet angularity $\tau_{a}$ (see definition in \cite{jet_angularity}) is used to study the inner structure of light-flavor jets and charm-jets. The right panel of Figure~\ref{fig:eic_jet} presents the comparison of jet angularity distributions for light-flavor jets and charm-jets with different power order $a$ values in $e+p$ collisions at $\sqrt{s}$ = 63.2 GeV and ratio between the two of their normalized jet angularity distributions. These results indicate that charm-jets have relatively broader jet shapes compared to light-flavor jets. Different flavor tagged jets may experience different nuclear modification effects, which will be a key to improve the knowledge of hadronization in medium.

\section{Summary and Outlook}
Initial simulation results have demonstrated good statistical precisions can be achieved for reconstructed heavy flavor hadron and jet measurements at the future EIC. These measurements can access the unique kinematic phase space to explore the nucleon/nuclei structure, provide enhanced sensitivities to hadronization processes in vacuum or medium and significantly reduce the extrapolated nuclear transport coefficient uncertainties. These physics studies will provide guidance for the ongoing EIC detector design and R$\&$D. The future EIC will reveal the mysteries of several unresolved questions through the heavy flavor and jet probes.


\end{document}